# The calculation of air filtration efficiency through the visual basic programming language

Giorgos Kouropoulos
etmecheng@gmail.com
Department of Energy Technology, Technological Educational Institution of Athens

**Abstract**

The present article describes the development of a software which was written in visual basic programming language. The software calculates the particle collection efficiency and penetration of a fibrous filter medium for given values of particle diameter, fiber diameter filter thickness, volume fraction etc, during the process of air filtration. The progress of the development of software is divided into two steps, the design of graphical user interface (GUI) and the code writing. The code is mainly based on the mathematical model of particle collection efficiency of fibrous filters media.

**Introduction**

During the process of air filtration, the filter medium retains solid microparticles that contained in the air. More specifically, fibers of the filter medium function as interceptors and block the motion of microparticles. The process of air filtration consists of three filtration mechanisms which occur simultaneously, diffusion, interception and inertial impaction mechanism.

The particles collection efficiency of a filter medium expresses the difficulty of microparticles to penetrate the medium while penetration expresses the facility of microparticles. If a filter medium has 100% particle collection efficiency then penetration is 0% and vice versa. The relationship between particle collection efficiency and penetration of the filter medium is defined by Eq. 1.

$$E = 1 - P \tag{1}$$

The penetration of a filter medium is calculated by Eq. 2 as a function of filter thickness, fiber diameter and other parameters related to filtration sub-mechanisms.

$$P = \exp\left(-\frac{4La_r}{\pi d_f}\sum n\right) \tag{2}$$

Where:

L: Thickness of filter medium (mm).

$d_f$: Fiber diameter of filter medium (mm).

$a_r$: The volume ratio between the total volume of fibers of filter $V_f$ to the total volume of filter medium $V_F$.

$\Sigma$n : The sum of parameters of each filtration mechanism (dimensionless).

The sum Σn is equal to:

$$\sum n = n_D + n_R + n_I \tag{3}$$

Where:

$n_D$: Dimensionless parameter of diffusion mechanism.

$n_R$: Dimensionless parameter of interception mechanism.

$n_I$: Dimensionless parameter of inertial impaction mechanism.

The $n_D$ dimensionless parameter is equal to:

$$n_D = 1.61 \left(\frac{1 - a_r}{Ku}\right)^{\frac{1}{3}} Pe^{-\frac{2}{3}} \tag{4}$$

Where:

Pe: Dimensionless Peclet number.

Ku: Dimensionless Kuwabara hydrodynamic factor.

Kuwabara hydrodynamic factor is equal to:

$$Ku = \frac{4a_r - a_r^2 - 3}{4} - \frac{ln a_r}{2} \tag{5}$$

The Peclet number is equal to:

$$Pe = \frac{3 \times 10^{-12} \pi \mu \upsilon d_f d_P}{kT \left[1 + \left(\frac{0.067}{d_P}\right)(2.492 + 0.84 \exp(-6.49 d_P))\right]} \tag{6}$$

Where:

υ: Velocity of air within the filter pipeline (m/sec).

μ: Absolute viscosity of water (Kg/m×sec).

k: Boltzmann constant (1.3708× $10^{-23}$J/$^{o}$K).

T: Absolute temperature of water ($^{o}$K).

The $n_R$ dimensionless parameter of interception mechanism is equal to:

$$n_R = \frac{(1 - a_r)N_R^2}{Ku(1 + N_R)} \quad (7)$$

Where:

$N_R$: Dimensionless factor. Is the ratio of particles diameter $d_P$(μm) to the average fiber diameter $d_f$(μm) of filter medium.

The $n_I$ dimensionless parameter of inertial impaction mechanism is equal to:

$$n_I = \frac{Stk \times J}{2Ku^2} \quad (8)$$

Where:

Stk: Stokes dimensionless number.
J: Dimensionless factor which is dependent by $N_R$ factor.

J factor is chosen according to:

For $N_R < 0.4$ J = $(29.6 - 28{a_r}^{0.62})N_R^2 - 27.5N_R^{2.8}$ (9)
For $N_R > 0.4$ J = 2 (10)

The Stokes number is equal to:

$$Stk = \frac{\rho d_P^2 \upsilon C_D}{18\mu d_f} \quad (11)$$

Where:

$C_D$: The dimensionless friction coefficient. This coefficient is selected by Reynolds number of air flow in the filter. Coefficient $C_D = 0.44$ is chosen for this study [4, 7].

With replacement of equations (8), (7), (4) to (3) and (3) to (2) then results the final function P($d_P$) of the filter medium.

$$P = \exp\left[-\frac{4La_r}{\pi d_f}\left[1.61\left(\frac{1 - a_r}{Ku}\right)^{\frac{1}{3}} Pe^{-\frac{2}{3}} + \frac{(1 - a_r)N_R^2}{Ku(1 + N_R)} + \frac{Stk \times J}{2Ku^2}\right]\right] \quad (12)$$

In the case of the filters media with non cylindrical section, the diameter of filter should be replaced by equivalent diameter. The equivalent diameter of a filter $d_E$ is equal to:

$$d_E = \frac{4A}{\Pi} \quad (13)$$

Where:

A: Area of filter element.
Π: Perimeter of filter element.

## Development of the software

The first step of the development of software starts with design of graphical user interface of application. For these purposes, the necessary graphical elements for the implementation of software are selected. These elements include text boxes, labels, command button and forms. For more details see Table 1.

Table 1. The number of graphical elements

| Variables | Graphical elements |
| --- | --- |
| E, P, $n_D$, $n_R$, $n_I$, Re (exported data) | 6 labels |
| $d_P$, L, $d_f$, $a_r$, μ, υ, T, ρ (imported data) | 8 text boxes |
| Description of imported data | 8 labels |
| Grouping of imported/exported data | 2 forms |
| For final calculations | 1 command button |

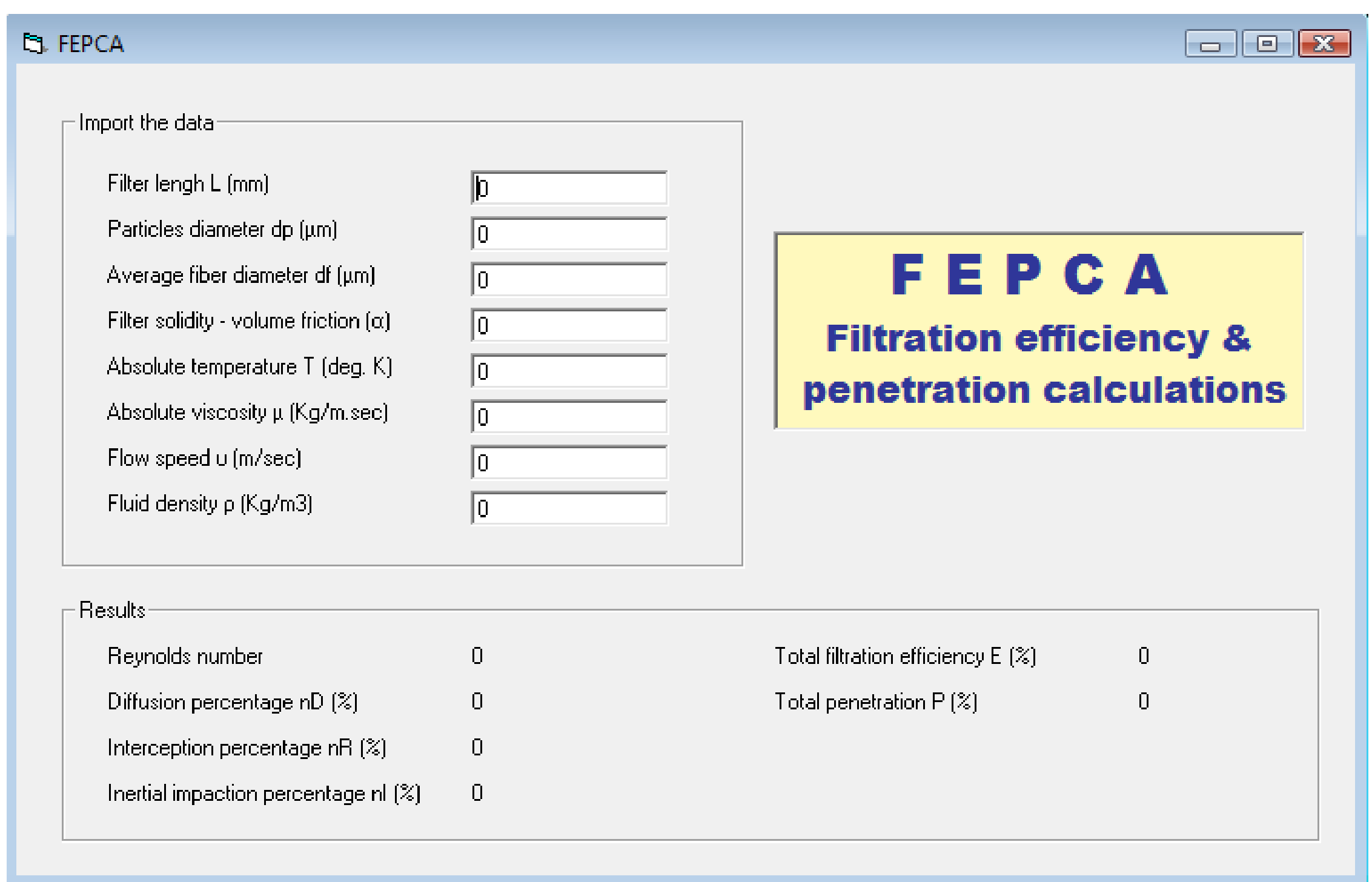

Fig. 1. Screenshot of FEPCA software (Filtration efficiency & penetration calculations)

## Writing of code

In this second step of the development progress of software, the mathematical model which calculates the particle collection efficiency for the filter medium (Eq. 1 - Eq. 12) will be converted into programming code in visual basic. Initially, it is necessary to become a correspondence between the physical quantities of mathematical model and the selected graphical elements of application. For more information see Table 2.

Table 2. Correspondence between physical quantities, variables and labels/textboxes of the graphical user interface

| **Physical quantity** | **Variable in the code** | **Labels** | **Text boxes** |
|---|---|---|---|
| L | L | Label1 | Text1 |
| $d_P$ | dp | Label2 | Text2 |
| $d_f$ | df | Label3 | Text3 |
| α | a | Label4 | Text4 |
| T | T | Label5 | Text5 |
| μ | m | Label6 | Text6 |
| υ | u | Label7 | Text7 |
| ρ | p | Label8 | Text8 |
| Re | Re | Label9 / Label10 | - |
| $n_D$ | nd | Label11 / Label12 | - |
| $n_R$ | nr | Label13 / Label14 | - |
| $n_I$ | ni | Label15 / Label16 | - |
| E | E | Label17 / Label18 | - |
| P | Pe | Label19 / Label20 | - |
| $d_F$ | df1 | Label21 | Text9 |
| Ku | ku | - | - |
| Pe | pe | - | - |
| $N_R$ | nr2 | - | - |
| Stk | stk | - | - |
| J | j | - | - |

The content of the programming code includes two basic parts, the declaration of variables for imported and exported data on the software according to Tables 1 - 2 and the purely mathematical part of code according to mathematical model. The declaration of variables follows below:

```
Dim L As Double
Dim dp As Double
Dim df As Double
Dim a As Double
Dim T As Double
Dim m As Double
Dim u As Double
Dim p As Double
Dim Re As Double
Dim nd As Double
Dim nr As Double
Dim ni As Double
Dim E As Double
Dim Pe As Double
```

The correspondence between imported data and text boxes in the code.

```
Private Sub Command1_Click()
L = Text1.Text
dp = Text2.Text
df = Text3.Text
a = Text4.Text
T = Text5.Text
m = Text6.Text
u = Text7.Text
p = Text8.Text

End Sub
```

The mathematical part of the code follows below:

```
Re = ((u * 10 ^ 6) * (df1 * 10 ^ 6) * (p / 10 ^ 18)) / (m / 10 ^ 6)
Label10.Caption = Re
ku = (((4 * a) - (a ^ 2) - 3) / 4) - Log(a) / 2
ch = 1 + (0.067 / dp) * (2.492 + (0.84 * Exp(-6.49 * dp)))
k = 1.3708 * 10 ^ -23
Pe = (3.14158 * (3 * 10 ^ -12) * df * dp * u * m) / (k * T * ch)
nd = 1.61 * (((1 - a) / ku) ^ (1 / 3)) * (Pe ^ (-2 / 3))
Label12.Caption = nd
nr2 = dp / df
nr = ((1 - a) * (nr2 ^ 2)) / (ku * (1 + nr2))
Label14.Caption = nr
stk = ((p / 10 ^ 18) * (u * 10 ^ 6) * (dp ^ 2)) / (40.91 * (m / 10 ^ 6) * df)
If nr2 < 0.4 Then
j = ((29.6 - 28 * (a ^ 0.62)) * (nr2 ^ 2)) - 27.5 * (nr2 ^ 2.8)
ElseIf nr2 > 0.4 Then
j = 2
End If
ni = (stk * j) / (2 * (ku ^ 2))
Label16.Caption = ni
Pe = 100 * Exp((-4 * a * (L * 10 ^ 3) / (3.14 * df)) * (nd + nr + ni))
E = 100 - Pe
Label20.Caption = Pe
Label18.Caption = E
End Sub
```

## Results

Below follow five diagrams which show the change of the particle collection efficiency of a filter medium as a function of the particle diameter for various values of fiber diameter, volume ratio of medium, velocity of air flow and the temperature of filtered air. The particle collection efficiency and penetration of the fiber filter medium will be calculated by the software for various values and parameters of the following quantities/variables.

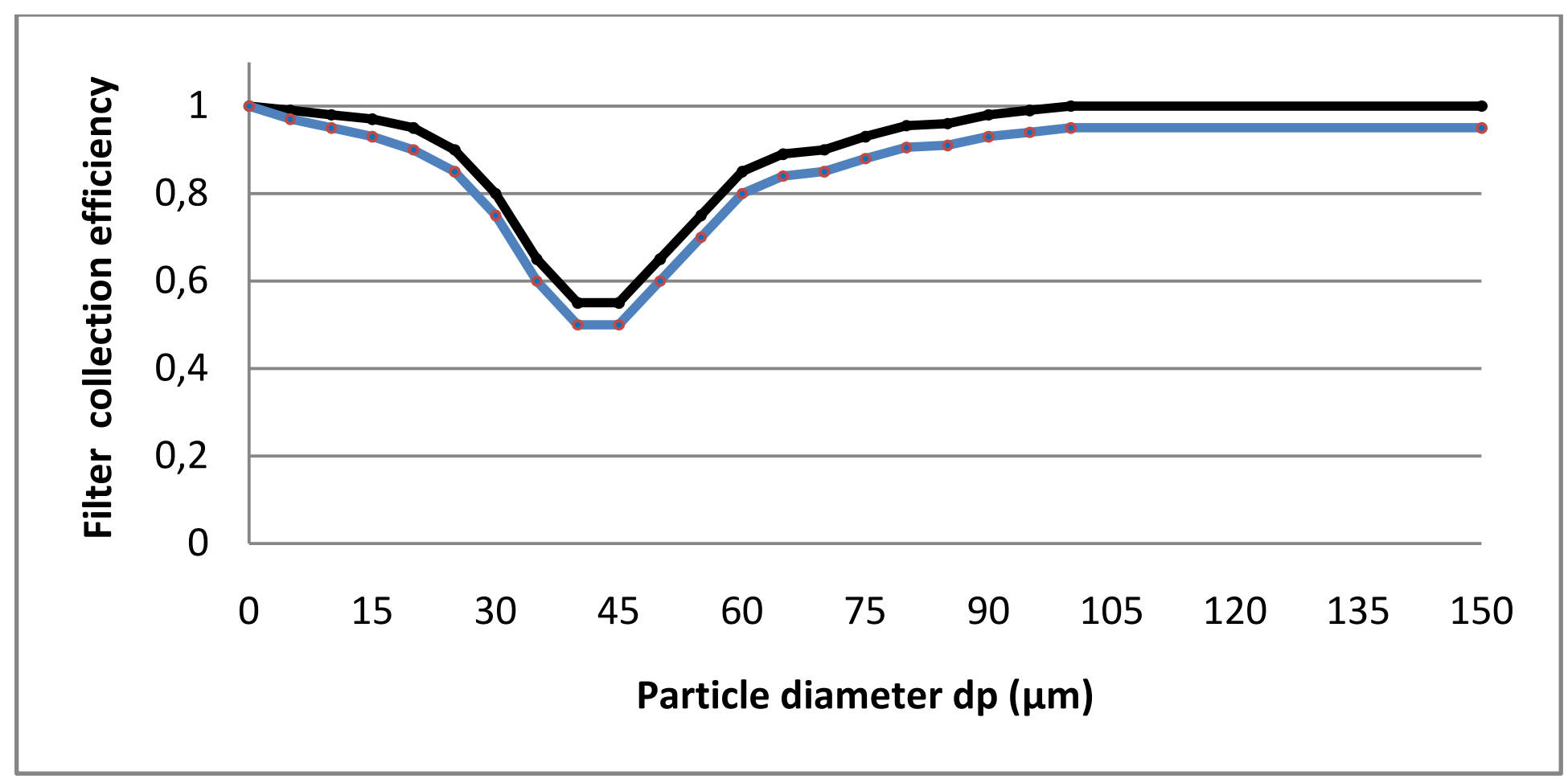

Fig. 2. The change of the filter collection efficiency in relation to particle diameter for air velocity 1 m/sec (black line) and 2.5 m/sec (blue line).

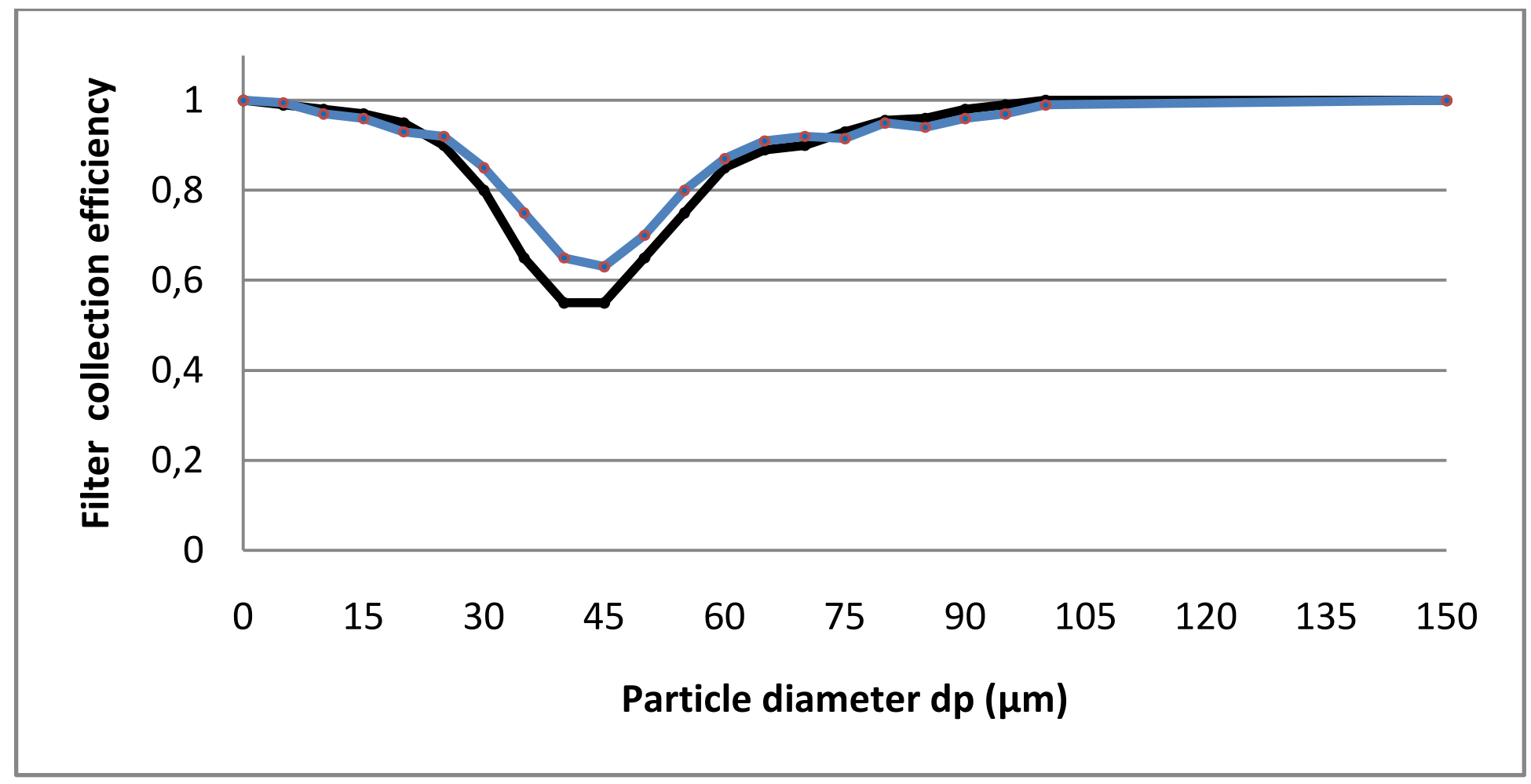

Fig. 3. The change of the filter collection efficiency in relation to particle diameter for fiber diameter 1µm (black line) and 4µm (blue line).

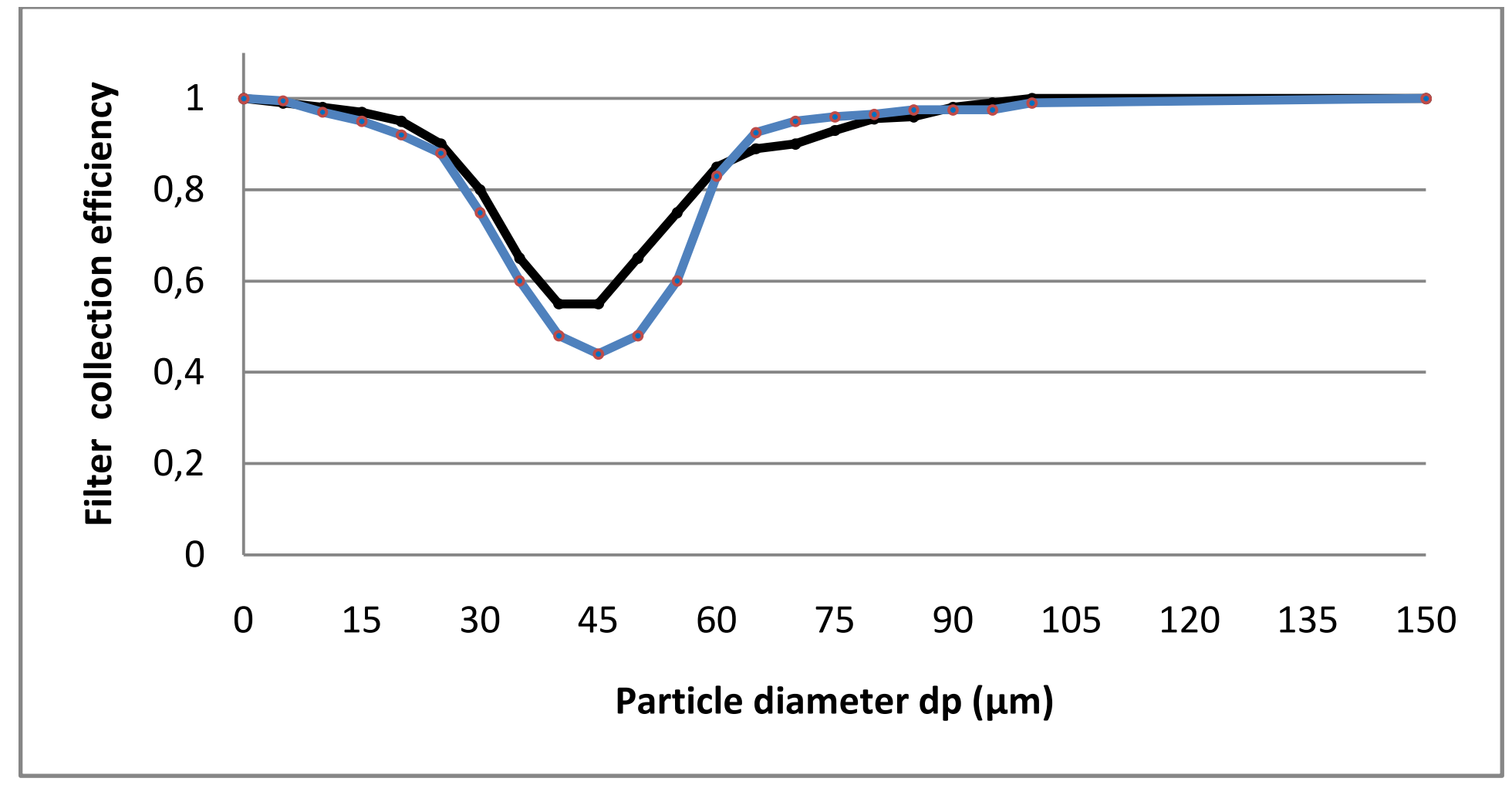

Fig. 4. The change of the filter collection efficiency in relation to volume ratio for volume ratio 0.95 (black line) and 0.5 (blue line).

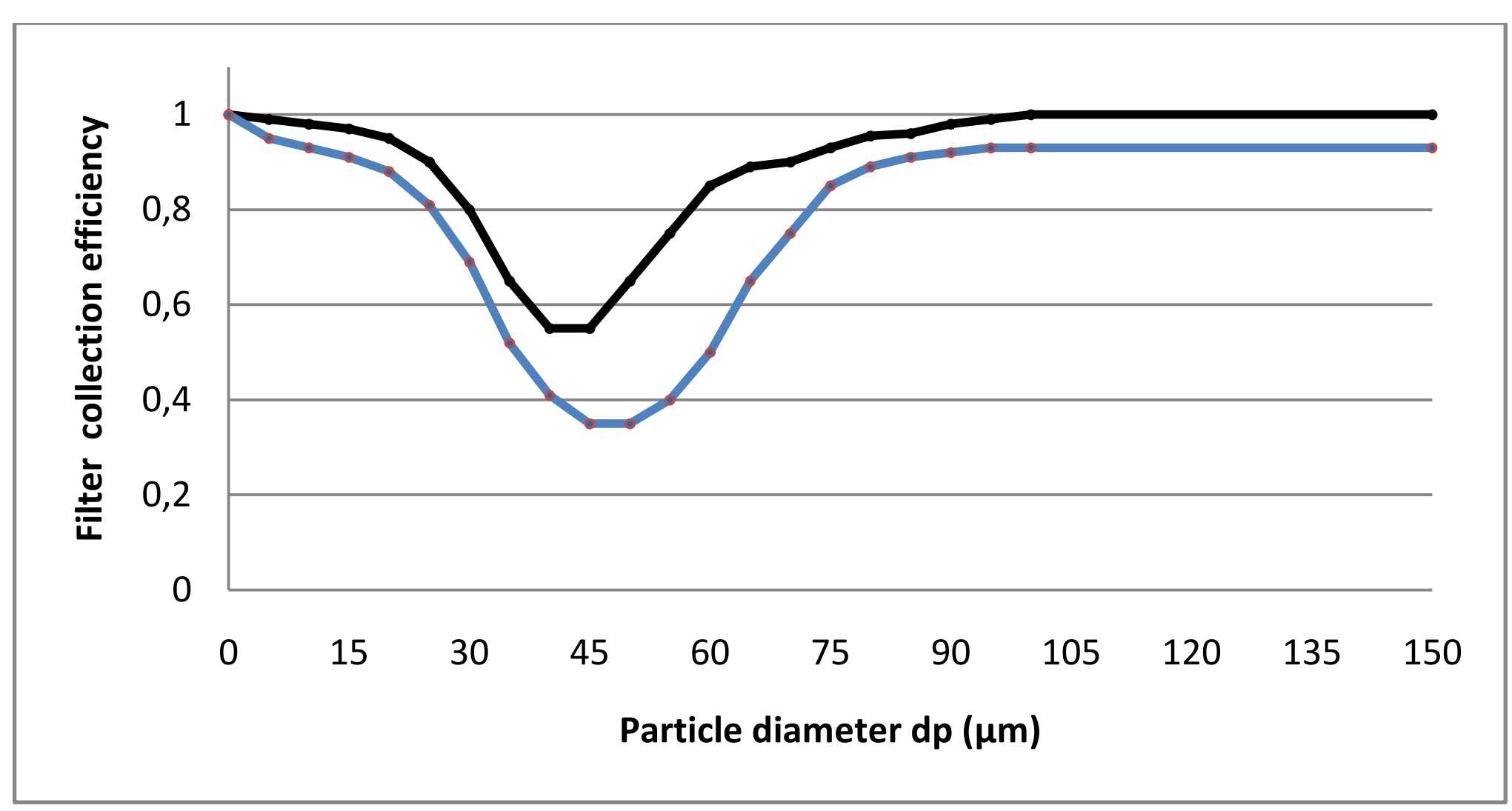

Fig. 5. The change of the filter collection efficiency in relation to temperature of filtered air for temperatures 20$^{o}$C (black line) and 50$^{o}$C (blue line).

## Conclusions

After the comparison of the results and exported data, in addition after studying of the figures and their relation to the theoretical curves of mathematical models, it is concluded that the software calculates accurately the particle collection efficiency of a fibrous filter medium during the process of air filtration. Consequently, the development of the software has been successfully completed. This software is suitable for educational, research and scientific purposes.